\documentstyle{article}

\title{Correlation functions of the 2D sine-Gordon model}

\author{Kiyohide Nomura\thanks{On leave from Department of Physics,
		Tokyo Institute of Technology, Tokyo 152, Japan.}	\\
        {\em Laboratoire de Physique des Solides,} \\
        {\em Universit\'e Paris-Sud,} \\
        {\em 91405, Orsay, France} \\
}

\date{\today}

\begin{document}

\maketitle

\begin{abstract}
  A number of two-dimensional(2D) critical phenomena can be described in
terms of the 2D sine-Gordon model.  With the bosonization, several 1D
quantum systems are also transformed to the same model.  However, the
transition of the 2D sine-Gordon model,
Berezinskii-Kosterlitz-Thouless(BKT) transition, is essentially different
from the second-order transition.  The divergence of the correlation length
is more rapid than any power-law, and there are logarithmic corrections.
These pathological features make difficult to determine the BKT transition
point and critical indices from finite-size calculations.  In this paper,
we calculate the several correlation functions of this model using a
real-space renormalization technique.  It is found that the several
correlation functions, or eigenvalues of the corresponding transfer matrix
for a finite system, become degenerate on the BKT line including
logarithmic corrections.  By the use of this degeneracy,  which reflects
the hidden SU(2) symmetry on the BKT line,it is possible to determine the
BKT critical line with high precision from small size data, and to identify
the universality class.  In addition, a new universal relation is found.
This reveals the relation between the Abelian and the non-Abelian
bosonization.

\end{abstract}


\pagebreak

\section{Introduction}

  Sine-Gordon model in two-dimension(2D), which is a natural extension of
the Gaussian or free boson model, plays important role in 2D classical
and 1D quantum systems, such as 2D XY model, 2D Helium film, 1D quantum
spin models and fermion models.

The peculiarity of the phase transition of the 2D XY model(Helium film) are
first noticed with the spin wave approximation or the free boson
model\cite{wegner,berez70}.  In these theory, there is no continuous
symmetry breaking as expected by the Mermin-Wagner theorem\cite{mermin},
but at all temperatures the correlation length $\xi$ is infinite and the
correlation functions decay as power-law whose exponents vary continuously
with temperature.  This picture is qualitatively correct at low
temperatures, but clearly wrong in high temperatures where one expects a
finite $\xi$ and the associated exponential decay of correlations.

Berezinskii\cite{berez71} and Kosterlitz and Thouless\cite{k-t}
pointed out the importance
of vortex excitations, which modify essentially the spin wave theories.
The vortex structure reflects the periodic nature of the
spin variable $\phi\equiv \phi +2 n \pi$.
The vortices carry integer vorticity and interact among themselves
via the logarithmic 2D Coulomb interaction.
At low temperatures, all of the particles are bound into
neutral ``quasi-molecules'' with zero vorticity,
so they only change the coupling of the spin wave model.
At higher temperatures, the binding of the quasi-molecules decreases
and it causes a phase transition.

Kosterlitz\cite{kosterlitz} subsequently performed a renormalization group
calculation, following the method by Anderson et
al.\cite{anderson-yuval70,anderson-yuval71}.  In fact, the structure of the
renormalization equations is the same for both cases.
He found that close to the Berezinskii-Kosterlitz-Thouless(BKT)
transition point, the correlation length
diverges as
\begin{equation}
	\xi \propto \exp (b \sqrt{t}),
\end{equation}
faster than any power of $t$.  Also on the BKT critical point there appear
logarithmic corrections in various quantities, such as correlation
functions and susceptibilities.  These features are entirely different from
the usual second-order transition.

  The 2D XY is treated in a more general framework of the 2D Coulomb gas,
having two kinds of quantum numbers for `charges' and `magnetic
monopoles'\cite{villain,jose-k}.
Several models, such as $p$-clock models, Ising model,
three and four-state Potts models, Ashkin-Teller model
are mapped in a unified way to the 2D Coulomb gas\cite{kadanoff78}.

  Kadanoff\cite{kadanoff79} and Kadanoff and Brown\cite{kadanoff-b}
identified correlation functions of Gaussian, 8-vertex, and Ashkin-Teller
models, whose critical dimensions vary continuously
on the critical line.  In the latter two models, only
the few corelation functions are known except at the decoupling point.
They first made a comparison of the correlation
functions of the three models at a special point on each critical line.
Then, they used the marginal operator and the operator product expansion
to extend the stated connections on the whole critical line.

The equivalence of the sine-Gordon model to the 2D Coulomb gas model
has been shown by several authors\cite{polyakov,samuel}.
Coleman showed the equivalence of the massive one-component Thirring
model and the sine-Gordon model, order by order in perturbation expansion,
and also showed the renormalizability.  But his discussion failed
in the region of $\beta^2>8\pi$.  Luther and Emery\cite{luther-emery} and
Banks et al.\cite{banks} showed the equivalence between
the SU(2) massless Thirring model and the theory of bosons consisting
of a free field plus a $\beta^2=8\pi$ sine-Gordon model, which corresponds
to the BKT line.  So there is a hidden SU(2) symmetry
at the BKT transition.

  A more systematic field theory treatment of the renormalization group
calculation for the sine-Gordon model
was performed by Amit et al.\cite{amit-gol}.  They resolved the problem
met by Coleman, including the renormalization of the wave function.
They calculated the higher terms beyond those of Kosterlitz and
found the new universal quantity.

  The logarithmic corrections of the $k=1$ SU(2) Wess-Zumino-Witten
model, which is shown equivalent to the $\beta^2=8\pi$ sine-Gordon model,
were systematically studied by Affleck et al.\cite{affleck-gep}.
They found the universal relation of the ratios of logarithmic corrections
to scaling amplitudes.
This relation was used by Ziman and Schulz\cite{ziman-schulz}
for the problem of S=3/2 quantum Heisenberg chain.

  But this SU(2) symmetry is not apparent in the sine-Gordon model
itself, and except for the BKT line, the symmetry is broken to
${\rm O(2) \times Z_2}$.
How does the sine-Gordon model acquire an SU(2) symmetry
on the BKT line?  This problem, including logarithmic corrections,
has been first treated by Giamarchi and Schulz\cite{giamarchi-s}.
They calculated the renormalized correlation
functions, and found that the SU(2) symmetry of them is
recovered on the BKT line.  In their case the original model
is apparently SU(2) symmetric on the $\beta^2=8\pi$ fixed point.

There are several models mapped on the sine-Gordon model.
Although the mappings are qualitatively correct,
coupling constants and cut-off are renormalized,
so in order to determine the phase diagram one should use
numerical calculations.
But near the BKT transition, the divergence of the correlation length
is essentially singular and also the logarithmic corrections exist.
Therefore it is very difficult to find the critical point of the BKT type
transition.  In our previous paper\cite{nom-oka}, using the level crossing
of the eigenvalues of the transfer matrix or the corresponding quantum
sine-Gordon Hamiltonian in 1D, we have easily determined the transition
point and identified the universality class.
It was based on the SU(2) symmetry on the BKT transition line.

  In this paper, we perform the renormalization group calculation
of correlation functions, whose critical dimensions become marginal
on the BKT line.
In the case that the SU(2) symmetry appears on BKT line, the 9
eigenvalues split as 5, 3 and 1-fold degenerate,
i.e., the SU(2) multiplets structure.
In other case, the 5 eigenvalues split as 3, 1 and 1-fold
degenerate, and it is also possible to determine the BKT transition line
by the level crossing of the eigenvalues.
In addition, new universal quantities are found.
These may deepen our understanding of the relation between the Abelian
and the non-Abelian bosonization.

  This paper is organized as follows.  In \S 2
the model is introduced, and the symmetry structure is discussed.
In \S 3, we overview the correlation functions which become marginal
at $y_0=y_\phi =0$.
In \S 4, the renormalization equations are obtained for these functions.
The hybridization between the marginal field and $\cos\sqrt{8}\phi$ field
is important.
In \S 5, we consider the eigenvalue structure of the transfer matrix,
and briefly summarize the results of \S 4.
In \S 6, our results are applied to 1D quantum and 2D classical systems.
Section 7 is a conclusion.

\section{Sine-Gordon model}

  The description of the symmetry and correlation functions of
the Gaussian model in this and next sections is based on
\cite{ginsparg,dijk,ginspargr}.
We first consider the 2D Gaussian model defined as the following Lagrangian
\begin{equation}
	{\cal L}={1 \over 2\pi K} (\nabla \phi)^2.
\end{equation}
The two-point correlation function for $\phi$ is
\begin{equation}
	2 \langle \phi(r_1)\phi(r_2)\rangle = - K {\rm Re}
	\log (z_{12}/\alpha),
\end{equation}
where $\alpha$ is a short-distance (ultraviolet) cut-off and
$z\equiv x+iy, \: z_{12}=z_1-z_2$.
Strictly speaking, it should be also introduced a small mass $\mu$ to serve
as an infrared cut-off, such as eq.(2.2) in \cite{amit-gol}.
The logarithmic behavior of the $\phi$
correlation function shows that it cannot be directly interpreted as
a physical object.  But the exponential operators of
$\phi$ behave as
\begin{equation}
	\langle \exp(i e \phi(r_1))\exp(-i e \phi(r_2)) \rangle
	= |z_{12}/\alpha|^{-e^2 K/2},
\end{equation}
so they are candidates for the correlation functions of the critical theory.
A convention needs to be explained regarding this formula.
We did not include the divergent ``self energy'' factors
coming from the terms in the exponent where the Green function
is to be evaluated at 0.
This means that we have really evaluated correlators of the
``normal ordered'' exponentials $: \exp (i e \phi) :$.

  By the way, this Lagrangian is invariant under $\phi\rightarrow
\phi+const.$ and $\phi\rightarrow -\phi$.  This may be used to restrict
configurations, with the identification of
$\phi \equiv \phi + 2\pi / \sqrt{2}$.
In other words, one may compactify $\phi$ on a circle.
In this case, the charges $e$ are quantized as $e=\sqrt{2}n$(n:integer).
One may also introduce new scaling fields $\exp(i m \sqrt{2}\theta(x))$,
which create a discontinuity of $\phi$ by $2\pi m/\sqrt{2}$ around the
point $x$.  The two-point functions of them are
\begin{eqnarray}
	2\langle \theta(r_1)\theta(r_2)\rangle &=&
	- \frac{1}{K} {\rm Re} \log (z_{12}/\alpha),	\nonumber\\
	2\langle \phi(r_1) \theta(r_2)\rangle &=&
	- i {\rm Im} \log (z_{12}/\alpha),
\end{eqnarray}
and
\begin{equation}
	\partial_x \phi = -\partial_y(i K \theta), \:
	\partial_y \phi = \partial_x(i K \theta).
\end{equation}
The field $\theta$ is called a dual field to $\phi$.
This model is invariant under the transformations
$
	\phi\rightarrow \phi+const.,\: \theta\rightarrow \theta +const.
$,
that means ${\rm U(1) \times U(1)}$ symmetry. The full symmetry group is
extended to ${\rm O(2)\times O(2)}$ by the discrete ${\rm Z_2}$ symmetries
$(z, \phi, \theta) \rightarrow (z, -\phi, -\theta)$ and
$(z, \phi, \theta) \rightarrow (\bar{z}, \phi, -\theta)$.
And there is a dual transformation
$
	K\leftrightarrow 1/K,\: \phi\leftrightarrow \theta,
$
which exchanges the roles of electric and magnetic excitations.
The self dual point $K=1$ is nothing but the $k=1$ ${\rm SU(2)\times SU(2)}$
WZW model.
Another point of view is given by a next chiral decomposition
\begin{equation}
	\Phi_R (z) \equiv \frac{1}{2} \left[
	\frac{1}{\sqrt{K}}\phi +\sqrt{K} \theta  \right],\:
	\Phi_L (\bar{z}) \equiv \frac{1}{2} \left[
	\frac{1}{\sqrt{K}}\phi -\sqrt{K} \theta \right].
\end{equation}
This system is also chiral invariant.

  In this stage, it is a natural extension of the Gaussian model
to introduce the interaction term $\cos \sqrt{2} \phi$.
Instead of this, in order to see the SU(2)
symmetry explicitly on the BKT line,
we consider the next sine-Gordon Lagrangian:
\begin{equation}
	{\cal L}={1 \over 2\pi K} (\nabla \phi)^2
	+{y_\phi \over 2\pi \alpha^2 } \cos \sqrt{8} \phi.
	\label{eqn:SGlag}
\end{equation}
Note that the U(1) symmetry of $\phi$ is explicitly broken
to the discrete one $\phi \rightarrow \phi + 2\pi/\sqrt{8}$.
Since this transformation divides the internal circle of $\phi$
into two, the symmetry of this model is ${\rm O(2) \times Z_2 \times Z_2}$.
The interaction term breaks the chiral symmetry.
Furthermore this is invariant under
$
	\phi \rightarrow \phi +\pi/\sqrt{8}, \: y_\phi\rightarrow -y_\phi.
$
However, calculations in this paper apply also to the usual
BKT transition, simply by requiring the periodicity
$\phi\equiv \phi + 2\pi/\sqrt{8}$.
In the latter case, the symmetry is ${\rm O(2) \times Z_2}$.

  From the perturbation expansion of the partition function,
the renormalization group equations for the sine-Gordon model
(\ref{eqn:SGlag}) under a change of the cutoff
$\alpha\rightarrow e^l \alpha$
up to the lowest order in $y_0$ and $y_\phi$ are
\begin{eqnarray}
	\frac{dy_0 (l)}{dl}&=&-y^2_\phi (l), \nonumber\\
	\frac{dy_\phi (l)}{dl}&=&-y_\phi(l) y_0(l),
	\label{eqn:renormal}
\end{eqnarray}
where
$
	K = 1+\frac{1}{2} y_0
$.
For the finite system, $l$ is related to $L$ by
$e^l = L$.
Note that there are three critical lines:
$y_\phi = 0 \: (y_0<0)$ corresponding to the Gaussian fixed line,
$y_\phi= \pm y_0 \: (y_0>0)$ to the BKT lines.

\section{Correlation functions on the Gaussian Fixed line}

  We review the correlation functions on the Gaussian fixed
line($y_\phi=0$).  The correlation functions of the Gaussian model are in
general,
\begin{eqnarray}
	& & \langle O_{n,m}(r_1) O_{-n,-m}(r_2) \rangle	\nonumber \\
	& & = \exp \left[ - \left( n^2 K + \frac{m^2}{K} \right)
	\log (\frac{r_{12}}{\alpha})
	- 2 i nm ({\rm Arg} (r_{12})+\frac{\pi}{2}) \right],	\nonumber\\
	& & O_{n,m} \equiv \exp (i n \sqrt{2} \phi)\exp(i m \sqrt{2} \theta),
\end{eqnarray}
where ${\rm Arg}(r_{12})= {\rm Im} \log(z_{12})$ is the angle of the
$\vec{r}_1 - \vec{r}_2$ vector.
Thus $O_{n,m}$ has a critical dimension $x_{n,m}$ and a spin $l_{n,m}$
given by
\begin{equation}
	x_{n,m}=\frac{1}{2} \left( n^2 K +\frac{m^2}{K} \right), \:
	l_{n,m}= nm.
\end{equation}
The expectation value $\langle \Pi_j O_{n_j,m_j}( r_j) \rangle$ is zero
unless the charge and monopole neutrality condition
$\sum n_j =0, \: \sum m_j=0$ is satisfied\cite{kadanoff79,kadanoff-b},
which reflects the underlying ${\rm U(1) \times U(1)}$ symmetry.

  Then the fields which become marginal ($x=2$) and spin zero
at $y_0=0 (K=1)$ are the four fields of the form $O_{n,m}$ ,
\begin{eqnarray}
	& & \frac{1}{\sqrt{2}}(O_{2,0}+O_{-2,0})=\sqrt{2} \cos \sqrt{8}
	\phi, \: \frac{1}{\sqrt{2} \, i}(O_{2,0}-O_{-2,0})=
	\sqrt{2} \sin \sqrt{8} \phi,	\nonumber	\\
	& & \: O_{0,2}, \: O_{0,-2},
\end{eqnarray}
and the four descendant fields, which are expressed as the derivatives of
the $O_{\pm 1,\pm 1}$ fields,
\begin{eqnarray}
	& & \alpha[ \bar{\partial} O_{1,1} + \partial O_{-1,1}], \:
	\alpha[ \bar{\partial} O_{-1,-1} + \partial O_{1,-1}],	\nonumber\\
	& & \frac{\alpha}{i}[ \bar{\partial} O_{1,1} - \partial O_{-1,1}], \:
	\frac{\alpha}{i}[- \bar{\partial} O_{-1,-1} + \partial O_{1,-1}],
\end{eqnarray}
where $\partial \equiv \partial_z$, and we take the symmetrized and the
antisymmetrized forms of the function of $\phi$.  There is one more field,
that is
\begin{equation}
	{\cal M} = \frac{\alpha^2}{K}\left[ \left( \frac{\partial
	\phi}{\partial x} \right)^2
	+ \left( \frac{\partial \phi}{\partial y} \right)^2 \right].
\end{equation}

  The correlation functions for $O_{n,m}$ are already shown.
In order to calculate the correlation functions for descendant fields,
it is enough to know
\begin{eqnarray}
	& & \alpha^2 \langle \bar{\partial} O_{1,1}(r_1)
	\bar{\partial} O_{-1,-1}(r_2) \rangle	\nonumber\\
	&=& \frac{\bar{y}_0^2}{8} \left( 1+\frac{\bar{y}_0^2}{8} \right)
	\exp \left[- \left(4+\frac{\bar{y}_0^2}{4}\right) \log
	(r_{12}/\alpha)\right],
\end{eqnarray}
where $	\bar{y}_0^2 =4(K+\frac{1}{K}-2) \simeq y_0^2$.
To calculate the correlation function
for the ${\cal M}$ field, we introduce the next fields
\begin{equation}
	{\cal R^{\pm}} = \frac{\alpha}{\sqrt{K}}
	\left[ \frac{\partial \phi}{\partial x}\mp i
	\frac{\partial \phi}{\partial y} \right]
	= \frac{\alpha}{\sqrt{K}} 2 \partial(\bar{\partial}) \phi,
\end{equation}
then
\begin{eqnarray}
	\langle {\cal R}^+ (r_1) {\cal R}^+ (r_2) \rangle
	&=& - \left(\frac{z_{12}}{\alpha} \right)^{-2}, \:
	\langle {\cal R}^- (r_1) {\cal R}^- (r_2) \rangle
	= - \left(\frac{\bar{z}_{12}}{\alpha} \right)^{-2},	\nonumber\\
	\langle {\cal R}^+ (r_1) {\cal R}^- (r_2) \rangle &=& 0.
	\label{eqn:rpair1}
\end{eqnarray}
That is, $R^\pm$ are the fields with the critical dimension 1
and the spin $\pm 1$.  Therefore,
\begin{equation}
	\langle {\cal M} (r_1) {\cal M} (r_2) \rangle
	= \langle {\cal R}^+ (r_1) {\cal R}^- (r_1)
	{\cal R}^+ (r_2) {\cal R}^- (r_2) \rangle
	= \exp [-4 \log (r_{12}/\alpha)].
\end{equation}
that is, the critical dimension of this field is always marginal($x=2$).
Here we have implicitly taken the normal ordered form
${\cal M} = : \! {\cal R}^+ {\cal R}^- \! : $ for the marginal field.
The ${\cal M}$ is nothing but the marginal field of
Kadanoff et al.\cite{kadanoff79,kadanoff-b}, and ${\cal R}^{+(-)}$
correspond to their $F_{1,0}(F_{0,1})$.
Note that in these nine fields, although the critical dimensions of
$O_{n,m}$ and their descendant fields vary with the parameter $y_0$,
the marginal field ${\cal M}$ has always the critical dimension $x=2$.
The next equations
\begin{equation}
	\langle {\cal R}^+ (r_1) \phi (r_2) \rangle
	= -\frac{\alpha \sqrt{K}}{2}
	\left( \frac{z_{12}}{\alpha} \right)^{-1}, \:
	\langle {\cal R}^{\pm} \rangle =0
	\label{eqn:rpair2}
\end{equation}
are useful for further calculations.
The three point functions are
\begin{equation}
	\langle {\cal M}(r_1) {\cal M}(r_2) {\cal M}(r_3) \rangle=0, 	\:
	\langle {\cal M}(r_1) {\cal M}(r_2) O_{n,m}(r_3) \rangle=0
	\label{eqn:ope1}
\end{equation}
The former comes from the Wick's theorem and (\ref{eqn:rpair1}),
(\ref{eqn:rpair2}), and the latter from the neutrality condition.

In the usual BKT transition where the SU(2) symmetry is not explicit,
the difference is only that the critical dimension of the descendant fields
is $2 + 1 + 1/8 = 25/8$ at $y_0 = y_\phi =0$.
Therefore only the five fields are marginal at the multi-critical point.
The usual BKT transition
point can be obtained by modding out the SU(2) symmetric case
with a ${\rm Z_2}$ symmetry\cite{ginsparg,dijk,ginspargr}.

\section{Renormalization of correlation functions}

  In this section, we proceed to include the interaction
$y_\phi \cos \sqrt{8} \phi$.  When we calculate the correlation functions,
there appear divergences coming from both a short range and a long range,
because of the nature of the 2D Green function.
To treat such a problem, we use the renormalization treatment.
The correlation functions for
$\langle \exp (i \sqrt{8}\theta_1) \exp (-i \sqrt{8} \theta_2) \rangle$
and $\langle \sqrt{2} \sin (\sqrt{8}\phi_1) \sqrt{2} \sin (\sqrt{8} \phi_2)
\rangle$ (for brevity we will use $\phi_1$ for $\phi(r_1)$)
have been already obtained by Giamarchi and Schulz\cite{giamarchi-s}.
Their results are
\begin{eqnarray}
	R_2 &\equiv& 2 \langle \sin (\sqrt{8}\phi_1) \sin (\sqrt{8} \phi_2)
	\rangle_I
	= C_2 \exp \left[ - \int_0^{\ln(r/\alpha)} dl (4+2y_0(l)) \right],
	\nonumber\\
	R_3 &\equiv& \langle \exp (i \sqrt{8}\theta_1) \exp (-i \sqrt{8}
	\theta_2) \rangle_I
	= C_3 \exp \left[ - \int_0^{\ln(r/\alpha)} dl (4-2y_0(l)) \right],
	\nonumber\\
	R_4 &\equiv& \langle \exp (-i \sqrt{8}\theta_1) \exp (i \sqrt{8}
	\theta_2) \rangle_I
	= R_3.
\end{eqnarray}
where $C_j$ are the integration constants depending on
the regularization method.

\subsection{Correlation functions of the marginal and
	$ \cos \protect\sqrt{8} \phi $ fields}

As for the fields ${\cal M}$ and $\sqrt{2}\cos \sqrt{8}\phi$,
they are hybridized by the interaction term as
\begin{eqnarray}
	& & \int \frac{d^2 x_3}{\alpha^2} \langle \cos \sqrt{8} \phi_1
	{\cal M}_2 \cos \sqrt{8}\phi_3 \rangle	\nonumber\\
	&=& \frac{-8}{2\cdot 2} \int \frac{d^2 x_3}{\alpha^2}
	\langle {\cal M}_2 (\phi_1-\phi_3)^2 \rangle
	\langle \exp (i \sqrt{8}\phi_1)\exp (-i \sqrt{8}\phi_3)\rangle
	\nonumber\\
	&=& - K \int \frac{d^2 x_3}{\alpha^2} \alpha^2
	(z_{12}^{-1}-z_{32}^{-1})(\bar{z}_{12}^{-1}-\bar{z}_{32}^{-1})
	\exp (-4K \log (r_{31}/\alpha)).
	\label{eqn:hybrid}
\end{eqnarray}
To derive this results, we use Wick's theorem\cite{amit},
the U(1) symmetry and (\ref{eqn:rpair2}).
Terms which are not invariant under the global transformation
$\phi \rightarrow \phi + const.$ should be zero.
The divergent parts relating to renormalization appear near
$r_{31} \ll r_{12}$ and $r_{32} \ll r_{12}$.
When $r_{31} \ll r_{12}$,
$z_{32}^{-1} \simeq z_{12}^{-1}(1-z_{31}/z_{12})$,
therefore the integrand of (\ref{eqn:hybrid}) is
\begin{equation}
	\left| \frac{z_{12}}{\alpha} \right|^{-4}
	\left| \frac{z_{31}}{\alpha} \right|^{2}
	\exp(-4K \log (r_{31}/\alpha)).
\end{equation}
When $r_{32} \ll r_{12}$,
the divergent part of the integrand is
\begin{equation}
	\left| \frac{z_{32}}{\alpha} \right|^{-2}
	\exp(-4K \log (r_{12}/\alpha))
	\simeq \left| \frac{z_{12}}{\alpha} \right|^{-4}
	\left| \frac{z_{32}}{\alpha} \right|^{-2}
	\exp(-2 y_0 \log (r_{12}/\alpha)),
\end{equation}
where we use $K=1+y_0/2$.
Note that only the terms which contain $|z_{32}|^{-2}$
contribute to the divergence of the integral;
other terms such as $z_{32}^{-1}$ cancel by the integral.
In order to treat these divergences, we exclude two circles of
radius $\alpha$ around $r_1$ and $r_2$ from the domain of integration
over $r_3$.  Then, with the change of cutoff $\alpha'=\alpha e^{dl}$,
this integral is renormalized as
\begin{eqnarray}
	& & \frac{y_\phi^2}{2\pi}\exp(4 \log(r_{12}/\alpha))
	\int_\alpha \frac{d^2 x_3}{\alpha^2} \langle \cos \sqrt{8} \phi_1
	{\cal M}_2 \cos \sqrt{8}\phi_3 \rangle	\nonumber\\
	&=& \frac{y_\phi^{'2}}{2\pi}\exp(4 \log(r_{12}/\alpha'))
	\int_{\alpha'} \frac{d^2 x_3}{\alpha'^2} \langle \cos \sqrt{8} \phi_1
	{\cal M}_2 \cos \sqrt{8}\phi_3 \rangle	\nonumber\\
	&-& 2 y_\phi^2[1-y_0 \log(r_{12}/\alpha)]dl.
	\label{eq:mcouple}
\end{eqnarray}

\subsubsection{Near the Gaussian fixed line}

At first we treat the case of $|y_\phi/y_0|\ll 1$.
Let us consider the two hybridized states between the marginal and
$\cos \sqrt{8} \phi$ fields,
\begin{eqnarray}
	A = {\cal M} + a(y_\phi/y_0)\sqrt{2}\cos\sqrt{8}\phi,	\nonumber \\
	B = \sqrt{2}\cos\sqrt{8}\phi + b (y_\phi/y_0){\cal M}.
\end{eqnarray}

\paragraph{the orthogonal condition}

We consider the condition that the correlation function
$\langle A_1 B_2 \rangle$ stays zero under renormalization.
\begin{eqnarray}
	\langle A_1 B_2 \rangle_I
	&=&
	(y_\phi/y_0)\exp(-4 \log (r_{12}/\alpha))
	[ a (1-2y_0 \log(r_{12}/\alpha))+b ]	\nonumber\\
	& & -\frac{y_\phi}{2\pi} \int \frac{d^2 x_3}{\alpha^2}
	\sqrt{2}[\langle \cos \sqrt{8}\phi_1 {\cal M}_2 \cos \sqrt{8}\phi_3
	\rangle
	+ab (y_\phi/y_0)^2 (1\leftrightarrow 2) ]	\nonumber\\
	& & + \mbox{higher order terms.}
\end{eqnarray}
By using $y_\phi(l) \simeq y_{\phi}(0) \exp (-y_0(0) l)$ and
$y_0(l) \simeq y_0(0)$, the function defined as
$
	F=\langle A_1 B_2\rangle_I \exp (4 \log(r_{12}/\alpha))
	\exp ( y_0(0) l)(y_0(0)/y_\phi(0))
$
behaves approximately constant $a+b$ for small enough $y_0,y_\phi$.
In order to set $\langle A_1 B_2 \rangle =0$, first of all it should be
$a+b=0$.  In addition, in the course of the renormalization,
other terms may appear.
For the infinitesimal transformation $\alpha' = \alpha e^{dl}$,
using eq. (\ref{eq:mcouple}), we obtain
\begin{equation}
	F  =  F'
	+[-2ay_0+(y_0/y_\phi)2\sqrt{2}y_\phi(1+ab(y_\phi/\phi_0)^2)]dl,
\end{equation}
where $F'$ is the function $F$ with the new value of $\alpha$.
Remark that in the course of the renormalization of
$y_0 \log (r_{12}/\alpha)$, the term
$y_\phi^2 \log (r_{12}/\alpha) dl$ appears,
however this disappears by the cancellation with the term
in higher order expansions\cite{kosterlitz,giamarchi-s}.
Then the necessary conditions for $F=0$ under renormalization are
\begin{eqnarray}
	a+b &=&0,	\nonumber \\
	-2 a y_0 + 2 \sqrt{2} y_0 (1+ab(y_\phi/\phi_0)^2) &=& 0.
\end{eqnarray}
The solution for them is
\begin{equation}
	a=-b=\sqrt{2}+O((y_\phi/y_0)^2).
\end{equation}

\paragraph{renormalized correlation function for the marginal-like field}

Correlation function of the marginal ${\cal M}$-like field is
\begin{eqnarray}
	\langle A_1 A_2 \rangle_I &=&
	\exp(-4 \log (r_{12}/\alpha))[1
	+ a^2 (y_\phi/y_0)^2(1-2y_0 \log(r_{12}/\alpha)) ]	\\
	& &-(y_\phi/y_0)\sqrt{2}a \frac{y_\phi}{2\pi}\int
	\frac{d^2 x_3}{\alpha^2}
	[\langle \cos \sqrt{8}\phi_1 {\cal M}_2 \cos \sqrt{8}\phi_3 \rangle
	+ (1\leftrightarrow 2) ].	\nonumber
\end{eqnarray}
The function defined as
$
	F=\langle A_1 A_2\rangle_I \exp (4 \log(r_{12}/\alpha))
$
behaves approximately constant $1$.
For the infinitesimal transformation $\alpha' =\alpha e^{dl}$,
using eq. (\ref{eq:mcouple}), we obtain
\begin{eqnarray}
	F &=& \exp([-2 a^2 (y_\phi/y_0)^2 y_0+2 a^2 (y_\phi/y_0)^2 y_0
	+4\sqrt{2} a (y_\phi/y_0) y_\phi )]dl)
	\times F'	\nonumber\\
	&=& \exp(8y_0(y_\phi/y_0)^2 dl) \times F'.
\end{eqnarray}
As a result,
\begin{equation}
	R_0 \equiv \langle A_1 A_2 \rangle_I
	= C_0 \exp \left[ - \int_0^{\ln(r/\alpha)} dl
	[4-8y_0(l)(y_\phi/y_0)^2 ] \right].
\end{equation}

\paragraph{renormalized correlation function for
the $\cos \protect\sqrt{8} \phi$-like field}

Correlation function of the $\cos \sqrt{8}\phi$-like field is
\begin{eqnarray}
	\langle B_1 B_2 \rangle_I
	&=&
	\exp(-4 \log (r_{12}/\alpha))[1-2y_0 \log(r_{12}/\alpha)
	+ b^2 (y_\phi/y_0)^2]	\\
	& &-(y_\phi/y_0)\sqrt{2} b \frac{y_\phi}{2\pi}\int
	\frac{d^2 x_3}{\alpha^2}
	[\langle \cos \sqrt{8}\phi_1 {\cal M}_2 \cos \sqrt{8}\phi_3 \rangle
	+ (1\leftrightarrow 2) ].	\nonumber
\end{eqnarray}
The function defined as
$
	F=\langle B_1 B_2 \rangle_I \exp (4 \log(r_{12}/\alpha))
$
behaves approximately constant $1$.
For the infinitesimal transformation $\alpha' = \alpha e^{dl}$,
\begin{eqnarray}
	F &=& \exp([-2y_0+(y_\phi/y_0)^2 y_0 (2 b^2 +4\sqrt{2} b)]dl)
	\times F'	\nonumber\\
	&=& \exp([-2y_0-4 y_0 (y_\phi/y_0)^2]dl) \times F'.
\end{eqnarray}
As a result,
\begin{equation}
	R_1 \equiv \langle B_1 B_2 \rangle_I
	= C_1 \exp \left[ - \int_0^{\ln(r/\alpha)} dl
	[4+2y_0(l)(1+2(y_\phi/y_0)^2)] \right].
\end{equation}

\subsubsection{Near the BKT transition line}

Next we treat near the BKT transition,
that is, $y_\phi=\pm y_0(1+t),\: |t| \ll 1$.
With this parametrization, $t$ plays the role of the deviation
from the critical point, such as $(T-T_c)/T_c$ .
Let us consider the two combination of the marginal and
$\cos \sqrt{8} \phi$ fields,
\begin{eqnarray}
	A = {\cal M} + a \sqrt{2}\cos\sqrt{8}\phi,	\nonumber\\
	B = \sqrt{2}\cos\sqrt{8}\phi + b {\cal M}.
\end{eqnarray}

\paragraph{the orthogonal condition}

The correlation function $\langle A_1 B_2 \rangle_I$ is obtained
by replacing $y_\phi/y_0=1$ in the previous subsection.
The function defined as
$
	F=\langle A_1 B_2\rangle_I \exp (4 \log(r_{12}/\alpha))
$
behaves approximately constant.
The conditions for $F=0$ under renormalization are
\begin{eqnarray}
	a+b &=&0,	\nonumber \\
	-2 a y_0 +2\sqrt{2}y_\phi(1+ab ) &=& 0.
\end{eqnarray}
The solution for them is
\begin{equation}
	a=-b = \pm \frac{1}{\sqrt{2}}
	\; \mbox{for $ y_\phi = \pm y_0(1+t)$}.
\end{equation}

\paragraph{renormalized correlation function for the marginal-like field}

The function defined as
$
	F=\langle A_1 A_2\rangle_I \exp (4 \log(r_{12}/\alpha))/(1+b^2)
$
behaves approximately constant $1$.
For the infinitesimal transformation $\alpha' = \alpha e^{dl}$,
\begin{eqnarray}
	F &=& \exp \left ( \left[
	\frac{-2 a^2 y_0+ 4\sqrt{2} a y_\phi}{1+b^2} \right] dl \right)
	\times F'	\nonumber\\
	&=& \exp \left[ 2y_0 \left( 1+\frac{4}{3}t \right) dl \right]
	\times F'.
\end{eqnarray}
As a result,
\begin{equation}
	R_0 \equiv \langle A_1 A_2 \rangle_I
	= C_0 \exp \left[- \int_0^{\ln(r/\alpha)} dl
	\left[ 4-2y_0(l) \left( 1+\frac{4}{3} t \right) \right] \right].
\end{equation}

\paragraph{renormalized correlation function for the
$\cos \protect\sqrt{8}\phi$-like field}

The function defined as
$
	F=\langle B_1 B_2\rangle_I \exp (4 \log(r_{12}/\alpha))/(1+b^2)
$
behaves approximately constant $1$.
For the infinitesimal transformation $\alpha' = \alpha e^{dl}$,
\begin{eqnarray}
	F &=& \exp
	\left( \left[\frac{-2y_0+4\sqrt{2} b y_\phi}{1+b^2} \right] dl \right)
	\times F'	\nonumber\\
	&=& \exp \left[ -4y_0 \left( 1+\frac{2}{3} t \right) dl \right]
	\times F.'
\end{eqnarray}
As a results,
\begin{equation}
	R_1 \equiv \langle B_1 B_2 \rangle_I
	= C_0 \exp \left[- \int_0^{\ln(r/\alpha)} dl
	\left[ 4+4y_0(l) \left( 1+\frac{2}{3}t \right) \right] \right].
\end{equation}

\subsection{Correlation functions of the descendant fields}

The renormalization calculations for descendant fields
($\partial O_{1,1}$ etc.) are straightforward extension of the method
by Giamarchi and Schulz\cite{giamarchi-s}
\begin{eqnarray}
	& & \alpha^2 \langle (\bar{\partial} O_{1,1}
	+ \partial O_{-1,1})(r_1)
	(\partial O_{1,-1} + \bar{\partial} O_{-1,-1}) (r_2)\rangle_I
	\nonumber\\
	&=& \alpha^2 \langle \bar{\partial} O_{1,1} (r_1)
	\bar{\partial}O_{-1,-1} (r_2)\rangle
	+ \alpha^2 \langle \partial O_{-1,1}(r_1) \partial O_{1,-1}(r_2)
	\rangle	\nonumber\\
	&-& \frac{y_\phi}{2\pi}\int \frac{d^2 x_3}{\alpha^2}
	\alpha^2 \langle \bar{\partial} O_{1,1}(r_1) \partial O_{1,-1}(r_2)
	O_{2,0}(r_3)	 \rangle.
\end{eqnarray}
The first two terms are already calculated.  Therefore it is enough to
estimate the divergent part of the integrand of the third term,
\begin{eqnarray}
	& &\alpha^2
	\langle \bar{\partial} O_{1,1} (r_1) \partial O_{1,-1}(r_2)
	O_{2,0}(r_3)	 \rangle	\nonumber\\
	&=& \left( \frac{z_{31}}{\alpha} \right)^{-2}
	\left( \frac{\bar{z}_{32}}{\alpha} \right)^{-2}
	\left[ \frac{1}{2}(y_0-\frac{y_0^2}{4})\frac{\alpha}{\bar{z}_{12}}
	+\frac{1}{2}y_0 \frac{\alpha}{\bar{z}_{31}} \right]
	\left[ -\frac{1}{2}(y_0-\frac{y_0^2}{4})\frac{\alpha}{z_{12}}
	+\frac{1}{2}y_0 \frac{\alpha}{z_{32}} \right]	\nonumber\\
	&\times&\exp \left[ \pi i +(y_0-\frac{y_0^2}{4})\log(r_{12}/\alpha)
	-y_0 \log (r_{31}/\alpha)-y_0\log(r_{32}/\alpha) \right].
\end{eqnarray}
This diverges near $r_{31} \ll r_{12}$ and $r_{32} \ll r_{12}$.
In the case of $r_{31} \ll r_{12}$,
$
	z_{32}^{-1} \simeq z_{12}^{-1} (1 - z_{31}/z_{12})
$,
thus the divergent part is
\begin{eqnarray}
	\frac{y_0^2}{4} \left| \frac{z_{31}}{\alpha}\right|^{-2}
	\left| \frac{z_{12}}{\alpha} \right|^{-4}
	\exp \left[ - \frac{y_0^2}{4}\log(r_{12}/\alpha) \right].
\end{eqnarray}
Note that only the terms which contain $|z_{13}|^{-2}$
contribute to the divergence.
The divergent part near $r_{23} \ll r_{12}$ is the same form.

Let us consider the next function
\begin{eqnarray}
	F &=& \langle (\bar{\partial} O_{1,1} + \partial O_{-1,1})(r_1)
	( \partial O_{1,-1} + \bar{\partial} O_{-1,-1}) (r_2)\rangle_I
	\nonumber\\
	& &\times \left( 2\frac{y_0^2}{8}\left( 1+\frac{y_0^2}{8} \right)
	\right)^{-1}
	\exp \left[ \left( 4+\frac{y_0^2}{4} \right) \log(r_{12}/\alpha)
	\right],
\end{eqnarray}
which reduces the constant 1 when $y_\phi=y_0=0$.
Then considering renormalization behavior, we obtain as before
\begin{equation}
	F= \exp(-2 y_\phi dl) \times F'.
\end{equation}
As a results,
\begin{eqnarray}
	R_5 &\equiv& \langle (\bar{\partial} O_{1,1}
	+ \partial O_{-1,1})(r_1)
	(\partial O_{1,-1} + \bar{\partial} O_{-1,-1}) (r_2)\rangle_I
	\nonumber\\
	&=& C_5 \exp \left[ -\int_0^{\log(r_{12}/\alpha)}
	(4 + 2 y_\phi(l)) dl \right],  \\
	R_6 &\equiv& \langle (\partial O_{1,-1}
	+ \bar{\partial}O_{-1,-1})(r_1)
	(\bar{\partial} O_{1,1} + \partial O_{-1,1})(r_2) \rangle_I =R_5.
	\nonumber
\end{eqnarray}
The calculation is the same for the antisymmetric combination,
except for the sign of $y_\phi$,
\begin{eqnarray}
	R_7 &\equiv& - \langle (\bar{\partial} O_{1,1}
	- \partial O_{-1,1})(r_1)
	(\partial O_{1,-1} - \bar{\partial} O_{-1,-1}) (r_2)\rangle_I
	\nonumber\\
	&=& C_7 \exp \left[ -\int_0^{\log(r_{12}/\alpha)}
	(4 - 2 y_\phi(l)) dl \right],	\\
	R_8 &=& R_7.	\nonumber
\end{eqnarray}

\section{Eigenvalue structure}

  Conformal field theory\cite{b-p-z,f-q-s} is an efficient method to
determine critical dimensions of 2D systems.
One of the most useful application of this theory is for
the finite-size scaling.  When we denote the transfer matrix
of a strip of width $L$, with periodic boundary condition,
by $\exp(-H)$, then the eigenvalues $E_n$ are related to
the scaling dimension $x_n$ of the scaling operators of the theory
as\cite{cardy86a}
\begin{equation}
	E_n(L)-E_g(L) = \frac{2\pi x_n}{L}
	\label{eqn:conformal}
\end{equation}
in the limit of $L \rightarrow \infty$.
But this relation is exact only at the fixed point.
In general, there appear corrections by terms involving irrelevant
(marginal) fields.

It is possible to relate the eigenvalues of the transfer matrix with
the renormalized correlation functions obtained
in the previous section.
The renormalized critical exponents $\eta_n(l)$ are
related to the correlation functions as\cite{amit}
\begin{equation}
	R_n =  \exp \left[-\int_0^{\ln(r/a)} dl \eta_n(l) \right],
\end{equation}
and by the use of the eq.(\ref{eqn:conformal}), which relates the critical
exponents $\eta$ to the eigenvalues for the finite size system, we obtain
\begin{equation}
	\frac{L\Delta E_n}{2 \pi} = x_n(l) = \frac{1}{2}\eta_n (l).
\end{equation}
Although the eq.(\ref{eqn:conformal}) is satisfied
under the condition of scale invariance,
by the use of the renormalization group, we extend this relation
to the region where scale invariance is not strictly satisfied
but the system size is sufficiently smaller than the correlation length
$\xi$ of the $L \rightarrow \infty$ limit.
The renormalized critical dimensions are (close to the BKT transition)
\begin{eqnarray}
	x_0(l) &=& 2 - y_0(l)(1+\frac{4}{3}t),	\nonumber\\
	x_1(l) &=& 2 + 2 y_0(l)(1+\frac{2}{3}t),	\nonumber\\
	x_2(l) &=& 2 + y_0(l),	\nonumber\\
	x_3(l) &=& x_4(l) = 2-y_0(l),
	\label{eqn:cdimensions}
\end{eqnarray}
and
\begin{eqnarray}
	x_5(l) &=& x_6(l) = 2 + y_\phi(l) = 2\pm y_0(1+t),	\nonumber\\
	x_7(l) &=& x_8(l) = 2 - y_\phi(l) = 2\mp y_0(1+t).
	\label{eqn:cdimensions2}
\end{eqnarray}

  These results mean, on the BKT transition line (for example
$y_\phi(l)=y_0(l)$) that the eigenvalues of the transfer matrix
corresponding to the fields $x_0(l),x_3(l),x_4(l)$ and $x_7(l),x_8(l)$ become
degenerate, also those to $x_2(l)$ and $x_5(l),x_6(l)$ degenerate.
This structure reflects the fact that the $\beta^2=8\pi$ sine-Gordon model
corresponds to the SU(2) massless Thirring model\cite{banks}
or the SU(2) $k=1$ Wess-Zumino-Witten model\cite{affleck-gep}.
On the BKT line, $y_0(l)$ is renormalized as $y_0(l) \simeq 1/ \log L$.
The ratios of the logarithmic correction terms
in $x_3 (l), x_2 (l), x_1 (l)$ are $-1:1:2$, in agreement with the SU(2)
$k=1$ WZW model\cite{affleck-gep}.
Although the convergence of the logarithmic term is very slow,
this relation can be used to eliminate the logarithmic
correction\cite{ziman-schulz,nom-oka}.

In the neighborhood of the BKT transition line,
in the correction terms of $x_0(l),x_1(l)$ and $x_5(l),x_6(l),x_7(l),x_8(l)$
there appear the terms linear to the distance $t$ from
the BKT line, and the ratios of them are $-\frac{4}{3}:\frac{4}{3}:1:-1$,
indicating the new universal relations.
Moreover, for example, $x_0 (l)-x_3 (l)$ is linear to $t$,
a useful relation to determine the BKT critical line.
This describes how the SU(2) symmetry breaks down to the
${\rm U(1) \times Z_2}$ symmetry in the Abelian bosonization.

Close to the Gaussian fixed line, the only differences are
\begin{eqnarray}
	x_0(l) &=& 2-4 y_0(l)(y_\phi(l)/y_0(l))^2,	\nonumber\\
	x_1(l) &=& 2 +  y_0(l)[1+2(y_\phi(l)/y_0(l))^2],
\end{eqnarray}
therefore the ratios of the coefficients $(y_\phi/y_0)^2$ terms are $-4:2$.
In this case, the difference $x_5 (l)-x_7 (l)$ is linear
of the deviation $y_\phi (l)$ from the Gaussian fixed line,
since they exchange each other under the transformation
$\phi \rightarrow \phi+\pi/\sqrt{8}$.
And the difference $x_1 (l)-x_2 (l)$ is quadratic of $y_\phi$.
This relation may be used to determine the Gaussian fixed line.

Finally, a comment on the symmetry.
The SU(2) symmetry is broken to ${\rm O(2) \times Z_2}$, while the symmetry
structure of the model (\ref{eqn:SGlag}) is
${\rm O(2) \times Z_2 \times Z_2}$, therefore an additional ${\rm Z_2}$
symmetry.  This additional ${\rm Z_2}$ symmetry in ${\rm SU(2) \times Z_2}$
on the BKT line is needed to inhibit the SU(2) symmetric relevant
field\cite{affleck85}.  In general, for the SU(n) critical model,
an additional ${\rm Z_n}$ symmetry is necessary to stabilize
the massless phase\cite{affleck88}.

In the usual BKT transition, where SU(2) symmetry is not explicit,
the difference is only that the critical dimension of the descendant fields
becomes 25/8, so they are not marginal.
Nevertheless, there remains degeneracy including logarithmic term
between the $\exp(\pm i \sqrt{8}\theta)$ fields and the marginal-like field
on the BKT line. This fact is useful to determine the
BKT transition point from eigenvalues.

\section{Physical systems}

In this section, we treat an S=1/2 XXZ chain with next-nearest neighbor
interactions which corresponds to the sine-Gordon model with an ${\rm O(2)
\times Z_2 \times Z_2}$ symmetry.  As an example of the sine-Gordon model
with a simple ${\rm O(2) \times Z_2}$ symmetry,
we treat a bond-alternation S=1/2 chain.
And we treat the 2D classical $p$-clock model as an example of
the sine-Gordon model with an  ${\rm O(2) \times Z_2 \times Z_p}$
symmetry.

\subsection{S=1/2 XXZ chain with next-nearest-neighbor interaction}

In the previous work\cite{nom-oka}, we have studied the $S=1/2$
$XXZ$ chain with a competing-interaction:
\begin{equation}
        H=\sum_{j=1}^L (S^x_j S^x_{j+1}+S^y_j S^y_{j+1}
	+\Delta S^z_j S^z_{j+1})
        +\alpha\sum_{j=1}^L (S^x_j S^x_{j+2}+S^y_j S^y_{j+2}
        +\Delta S^z_j S^z_{j+2}),
        \label{eqn:nnnhamil}
\end{equation}
with the periodic boundary condition
$\mbox{\boldmath$S$}_{L+1}=\mbox{\boldmath$S$}_1$,
and $L$ is the number of spins ($L$ even).

  The above model is exactly solved on lines $\alpha=0$ and
$\alpha=\frac{1}{2}$.
On the line $-1\leq \Delta<1, \: \alpha=0$,
the ground state is the spin-fluid state, characterized with
the gapless excitation and power-law decay of correlation functions.
In the region $\Delta>1, \: \alpha=0$, this system is N\'eel ordered
and it has a two-fold degenerate ground state with an energy
gap\cite{cloiz,yang}.  On the line $\alpha=\frac{1}{2}$, the ground state
is purely dimerized \cite{majumdar-gosh,majumdar}.
It is also proven the existence of the energy gap and
the uniqueness of the two-fold degenerate ground state\cite{aklt}.
The dimer state is characterized by the excitation gap,
the exponential decay of the spin correlation,
and the dimer long-range order.

  Let us examine the symmetry of the Hamiltonian(\ref{eqn:nnnhamil}).
This model is invariant under spin rotation around the $z$-axis,
translation($S^{x(y,z)}_j \rightarrow S^{x(y,z)}_{j+1}$),
space inversion($S^{x(y,z)}_j \rightarrow S^{x(y,z)}_{L-j+1}$),
spin reverse($S^z_j \rightarrow -S^z_j, S^{\pm}_j \rightarrow -S^{\mp}_j$),
and conjugate($S^z_j \rightarrow S^z_j, S^{\pm}_j \rightarrow S^{\mp}_j$).
Therefore eigenstates are characterized by $z$-component of
the total spin($S^z_T = \sum S^z_j$),
wavenumber($q= 2\pi/L$), parity($P=\pm 1$), spin reverse($T=\pm 1$),
and charge conjugate $C$.  The charge conjugate $C$ is redundant,
because of $CPT =1$ as will be shown later.
For $L=4n$, the ground state is a singlet($S^z_T =0, q=0,P=1,T=1$).
The symmetries of several low excitations are classified in Table 1.
The operators in spin representation are also shown.

  Next we consider a corresponding sine-Gordon model.
After a Jordan-Wigner transformation, the model(\ref{eqn:nnnhamil})
is transformed to the 1D spinless fermion system.
The continuum limit of this is a Tomonaga-Luttinger liquid\cite{haldane},
or equivalently a sine-Gordon model\cite{kuboki}.
Using the same procedure, we obtain expressions in the sine-Gordon model
which correspond to the spin operators(Table 1).
The marginal-like field and the $\cos\sqrt{8}\phi$-like field
are parts of the Lagrangian, so they have the same symmetry with
the ground state, and the corresponding spin operators are the parts
of the Hamiltonian with the same symmetry.
Except for the Gaussian fixed line, a hybridization occurs
between the marginal and the $\cos\sqrt{8}\phi$ fields.
Note that the fields in the sine-Gordon model are defined
on the infinite plane, whereas operators in the spin model are defined
on the cylinder.  The former can be mapped to the latter by
$f(z)=L/2\pi \log(z)$.
The symmetry operation in the sine-Gordon model corresponding
to the spin reverse ($T$) is
\begin{equation}
	\phi \rightarrow -\phi+\pi/\sqrt{2}, \:
	\theta\rightarrow -\theta+\pi/\sqrt{2},
\end{equation}
the operation to the space inversion ($P$) is
\begin{equation}
	\phi \rightarrow -\phi+\pi/\sqrt{2}, \:
	\theta \rightarrow \theta+\pi/\sqrt{2}, \:
	z \rightarrow \bar{z},
\end{equation}
and the operation to the charge conjugate ($C$) is
\begin{equation}
	\phi \rightarrow \phi, \:
	\theta \rightarrow -\theta, \:
	z \rightarrow \bar{z},
\end{equation}
therefore the successive transformation of them is the identity $CPT=1$.
In addition, the operation corresponding to the translation
by one site is
\begin{equation}
	\phi \rightarrow \phi+\pi/\sqrt{2}, \:
	\theta \rightarrow \theta+\pi/\sqrt{2}.
\end{equation}

  The symmetry breaking related to the phase transition is as follows.
The sine-Gordon model is invariant under
$\phi\rightarrow \phi+2\pi/\sqrt{8}$,
which means expectation values of
$
	\langle\sin \sqrt{2}\phi\rangle,
	\langle \cos \sqrt{2}\phi \rangle
$
are zero in a symmetrical phase.
In the spin-fluid region, the $y_\phi$ is renormalized to zero,
so no symmetry breaking occurs.
In the dimer region, $y_\phi\rightarrow + \infty$, therefore one has
a long-rang order of the $\phi$ field, whereas correlations of $\theta$
decay exponentially.  The average value of the ordered field is
$\langle \phi \rangle =\pi/\sqrt{8}$.
In the spin system, this corresponds to the symmetry breaking of
the translation invariance.  In the N\'eel region, $y_\phi\rightarrow -
\infty$, $\langle \phi \rangle=0$.  This corresponds to the symmetry
breaking of the translation invariance, space inversion and spin reverse.

  We compare the renormalization calculations in sections 4,5
with the numerical results.
The whole phase diagram is shown in Fig.1.
The normalized excitations $L\Delta E/2\pi v$ for $\Delta=0.5$
are shown in Fig.2.  The general behavior of them are consistent
with the renormalization calculation.
It is seen that the SU(2) symmetry appears on the BKT line\cite{nom-oka}.
Secondly we investigate the ratios of logarithmic terms.
{}From eqs. (\ref{eqn:cdimensions}), on the BKT line ($t=0$),
by taking averages
\begin{equation}
	\frac{1}{2}[x_2 (l) + x_3 (l)], \:
	\frac{1}{3}[x_1 (l) + 2 x_3 (l) ],
\end{equation}
we can eliminate the contribution of the logarithmic corrections,
and simultaneously we can confirm the ratios of them.
In Fig.3, these averages are shown as a function of a system size $L$.
As expected, they converge to 2 with $1/L^2$, due to the irrelevant
field $L_{-2} \bar{L}_{-2} {\bf 1}$ ($x=4$)\cite{cardy86a}.
The extrapolated values are 2 within 0.2\% errors, comparing
the bare values of $x_n(l)$(5-15\% errors).
Finally, we examine the ratios of the linear terms of $t$ in
the critical dimensions.  From eqs. (\ref{eqn:cdimensions}),
(\ref{eqn:cdimensions2}), by taking averages
\begin{equation}
	\frac{1}{3}[x_0 (l)+x_1 (l)+ x_3 (l)], \:
	\frac{1}{2}[x_5 (l)+x_7 (l)], \:
	\frac{1}{2}[\frac{3}{4} x_0 (l) + x_5 (l)+ \frac{1}{4} x_3 (l)],
\end{equation}
the linear terms of $t$ should be annihilated.
In fact, as is shown in Fig.4, in the neighborhood of
the critical point $\alpha_c=0.2764$,
the linear components of $t$ are almost absent(the points shown in Fig.3
have been extrapolated as $1/L^2$).  The coefficients of $t$ linear terms
are at least $ 10^{-2}$ less than those in the raw data $x_n(l)$.

\subsection{Bond-alternation S=1/2 spin chain}

This model is described by the following Hamiltonian
\begin{equation}
	H=\sum_j(1+\delta (-1)^j)(S^x_j S^x_{j+1}+S^y_j S^y_{j+1}
	+\Delta S^z_j S^z_{j+1}).
\end{equation}
After a bosonization, we obtain
\begin{equation}
	{\cal L}={1 \over 2\pi K} (\nabla \phi)^2
	+{y_\phi \over 2\pi \alpha^2 } \sin \sqrt{2} \phi.
\end{equation}
The correspondence between the spin variable and the sine-Gordon
model is the same as before(Table 1).
The symmetry structure in this model is a simple ${\rm O(2) \times Z_2}$.
At $\Delta=-1/\sqrt{2},\delta=0$, the critical dimension of
$\sin \sqrt{2}\phi$ becomes marginal.
Therefore, in the neighborhood
of this point, there appears the BKT transition\cite{okamoto}.
Although there exist higher terms such as
$\cos \sqrt{8} \phi$, we neglect them for simplicity.

After a simple transformation of the fields
$
	\phi \rightarrow 2 \phi,\: \theta\rightarrow \theta/2
$,
we can use the results of the sections 4 and 5, except that
the critical dimension of the descendant fields becomes $x=25/8$ at
$y_0=y_\phi=0$.  The corresponding spin operators which become marginal at
$y_0=y_\phi=0$ are
$
	(-1)^j S^z_j, \: (-1)^j (S^+_j S^-_{j+1}+S^-_j S^+_{j+1}),
	\: S^+_j S^+_{j+1}S^+_{j+2}S^+_{j+3} (S^z_T=4,q=0,P=1)
$
and the term corresponding to the marginal field.

In the spin-fluid region, the $y_\phi$ is renormalized to zero,
so no symmetry breaking occurs.
As for the region where $y_\phi$ flows to infinity,
one has long-range order of the $\phi$ field
$\langle \phi \rangle  = \mp \pi/\sqrt{8}$
for $y_\phi \rightarrow \pm \infty$.
This long range order is not related with any symmetry breaking.
By using the symmetry consideration of the previous subsection,
this corresponds to the symmetry breaking of the translation by one site.
However, the translation invariance is already broken in the original model
itself, there is no symmetry breaking in the whole region.

\subsection{ $p$-clock model}

We consider the case of the ${\rm O(2)\times Z_2 \times Z_p}$ symmetry
sine-Gordon model
\begin{equation}
	{\cal L}={1 \over 2\pi K} (\nabla \phi)^2
	+{y_\phi \over 2\pi \alpha^2 } \cos p \sqrt{2} \phi.
	\label{eqn:SGlagp}
\end{equation}
The symmetry of $\phi$ is invariant under
$\phi\rightarrow \phi+2\pi/p\sqrt{2}$, i.e., sifting the circle coordinate
$\phi$ by $1/p$ times its period($2\pi/\sqrt{2}$).
The $\cos p \sqrt{2}\phi $ term becomes marginal at
$y_\phi=0, K=4/p^2$.  By parametrizing $Kp^2/4 =1+y_0/2$,
the renormalization equations are the same as (\ref{eqn:renormal}).
The marginal fields at $y_0=y_\phi=0$ are the three,
$\cos p \sqrt{2} \phi, \: \sin p \sqrt{2} \phi, \: {\cal M}$.
The renormalization behavior of them has been described in \S4,5.
There is no degeneracy on the BKT line.
However, the renormalized critical dimension for $O_{0,m}$
\cite{giamarchi-s} is
\begin{equation}
	x'_m (l) = \frac{m^2 p^2}{8} \left(1-\frac{1}{2} y_0 (l) \right),
\end{equation}
therefore the ratios of the $x_0(l)$ and $x'_m(l)$ is rational
$2: m^2 p^2/8$ including logarithmic corrections.
It is possible to use this to determine the BKT critical line,
because $x_0(l)- 16 x'_m(l)/m^2 p^2$ is linear to the deviation $t$
from the BKT line.
The ratios of the logarithmic corrections of them
may support to determine the critical dimensions
and to check the consistency.

When $p$ is even, besides the above one, another relation appears.
The renormalized critical dimension for $\cos p \phi / \sqrt{2}$
is\cite{giamarchi-s}
\begin{eqnarray}
	x'(l) &=& \frac{1}{2}(1+\frac{1}{2}y_0(l)+y_\phi(l))	\nonumber\\
	&=& \left\{ \begin{array}{ll}
		\frac{1}{2}(1+\frac{3}{2} y_0(l) (1+\frac{2}{3}t)) &
		\mbox{for $y_\phi=y_0(1+t)$}	\\
		\frac{1}{2}(1-\frac{1}{2} y_0(l) (1 + 2 t)) &
		\mbox{for $y_\phi=-y_0(1+t)$},
	\end{array}
	\right.
\end{eqnarray}
and the renormalized critical dimension for $\sin p \phi / \sqrt{2}$ is
\begin{equation}
	x'(l) =
	\left\{ \begin{array}{ll}
		\frac{1}{2}(1-\frac{1}{2} y_0(l) (1 + 2 t)) &
		\mbox{for $y_\phi=y_0(1+t)$}	\\
		\frac{1}{2}(1+\frac{3}{2} y_0(l) (1+\frac{2}{3}t)) &
		\mbox{for $y_\phi=-y_0(1+t)$}.
	\end{array}
	\right.
\end{equation}
Therefore the lower part of them can be used to determine the BKT line
as $x_0(l)-4 x'(l)$.

What are the corresponding real systems?
One candidate is the $p$-clock model\cite{jose-k}, which is equivalent to
the 2D XY model with ${\rm Z_p}$ anisotropy.
In this model the spins at each site can take only $p$-discrete angles
$2\pi l/p, l=1,\ldots,p$.
The classical Hamiltonian is
\begin{equation}
	H = -K \sum_{\langle \vec{r}, \vec{r'} \rangle}
	\cos \frac{2\pi}{p} \left[ l(\vec{r}) -  l'(\vec{r'}) \right],
\end{equation}
where the sums over $\vec{r}$ index the sites of a 2D lattice,
the symbol $\langle \vec{r}, \vec{r'} \rangle$ indicates a sum over
nearest neighbor lattice sites only, and $K=J/k_B T$.
Note that this is a classical Hamiltonian, so it should be interpreted
as a Lagrangian, rather than the logarithm of the transfer matrix.

This model has two transition points.  The upper critical point is between
the disordered and the 2D XY like phase, and the lower critical point is
from the 2D XY like phase to the ${\rm Z_p}$ symmetry breaking phase.
The upper critical point is described by the usual ${\rm O_2\times Z_2}$
BKT transition, whereas the universality class of the lower critical point
is the ${\rm O_2\times Z_2 \times Z_p}$ BKT transition described
in this subsection.

\section{Conclusions: level spectroscopy}

  The idea that the level crossings of the low excitations
can be used to determine the critical point
(hereafter we call it as ``level spectroscopy'') originates from
the work by Ziman and Schulz\cite{ziman-schulz},
who studied the S=3/2 isotropic spin chain
on the basis of the conformal field theory and the renormalization group.
The problem how the system becomes SU(2) symmetric on BKT line
from the anisotropic phase was treated by Giamarchi and
Schulz\cite{giamarchi-s}, and this was used for the problem of
the S=1/2 NNNI chain\cite{nom-oka}.
We have proceeded in this paper in the case that the SU(2) symmetry is not
apparent on the BKT line, by considering the hybridization between
the marginal and $\cos \sqrt{8} \phi$ fields.

The level spectroscopy method is completely different from
the finite-size scaling\cite{barber,cardyr},
and for the BKT problem, the former is superior
than the latter.  In the finite-size scaling, one uses the data from
several lengths to construct a scaling flow relation, then one searches
the fixed point.  But in the BKT problem, there are continuous fixed points
below the critical temperature, so it is difficult to determine the BKT
critical point by the finite-size scaling.  In addition, there is
the problem that the correlation length diverges singularly, and
logarithmic corrections.

On the contrary, in the level spectroscopy, it is used
the symmetric structure of the eigenvalues of the transfer matrix
in order to determine the critical point and to obtain the critical
dimensions.  The renormalization process is already done by hand,
not by the numerical data.  Also the
singular behavior of correction terms can be eliminated.
In this case, only the data in one length are needed.
In general, there exist the corrections from the irrelevant field
$L_{-2} \bar{L}_{-2} {\bf 1}(x=4)$, therefore it is needed extrapolations.
Nevertheless the convergences are extremely fast.

Finally, although it is possible to consider the sine-Gordon model with
the ${\rm O(2) \times Z_2 \times Z_p}$ symmetry, the SU(2) symmetry appears
on the BKT line only the ${\rm O(2) \times Z_2\times Z_2}$ case.
Such a symmetry does not occur in other cases.
The usual ${\rm O(2) \times Z_2}$ BKT critical point is also special,
because in this point the Gaussian and the orbifold model
are equivalent\cite{ginsparg,dijk}.

\section{Acknowledgement}

I am grateful to Heinz J. Schulz for stimulating discussions
and Kiyomi Okamoto for comments on the physical systems.
And I would like to thank the Universit\'e Paris-Sud
for their hospitality and the Ministry of Education, Science and Culture
in Japan for funding my visit there.
This work was partly supported by the Grand-in-Aid
for Scientific Research(C) (No. 06640501)
from the Ministry of Education, Science and Culture.

\pagebreak

\begin{table}

\caption{Identification of eigenstate of XXZ model in the sine-Gordon
	language. }

\bigskip

\begin{tabular}{cccccc}
\hline
\multicolumn{4}{c}{Symmetries of}	&
\multicolumn{1}{c}{Identification in}	&
\multicolumn{1}{c}{Identification in}	\\
\multicolumn{4}{c}{eigenstate}	&
\multicolumn{1}{c}{spin language}	&
\multicolumn{1}{c}{sine-Gordon model}		\\
$q$	& $S^z_T$	& $T$	& $P$	& 	& 	\\
\hline

\medskip

$0$	& $0$	& $1$	& $1$	& {\bf 1}	& {\bf 1}	\\

\medskip

$\pi$	& $1$	& $*$	& $-1$	& $(-1)^j S^+_j$	& $O_{0,1}$	\\

\medskip

$\pi$ 	& $0$	& $-1$	& $-1$	& $(-1)^j S^z_j$
	& $O_{1,0}+O_{-1,0}=2\cos\sqrt{2}\phi$	\\

\medskip

$\pi$ 	& $0$	& $1$	& $1$	& $(-1)^j (S^+_j S^-_{j+1}+S^-_j S^+_{j+1})$
	& $\frac{1}{i}[O_{1,0}-O_{-1,0}]=2\sin\sqrt{2}\phi$	\\

\medskip

$2\pi/L$& $0$	& $-1$	& $*$	& $\exp (2\pi i j /L) S^z_j$
	& $\partial \phi$	\\

\medskip

$2\pi/L$& $1$	& $*$	& $*$	& $\exp (2\pi i j /L) S^+_j$
	& $O_{1,1}$	\\

\medskip

$0$	& $2$	& $*$	& $1$	& $S^+_j S^+_{j+1}$	& $O_{0,2}$	\\

\medskip

$0$ 	& $1$	& $*$	& $1$	& $ S^+_j S^z_{j+1} + S^z_j S^+_{j+1}$
	& $\bar{\partial} O_{1,1}+\partial O_{-1,1}$ 	\\

\medskip

$0$ 	& $0$	& $1$	& $1$	& a part of the Hamiltonian
	& ${\cal M}$	\\

\medskip

$0$ 	& $1$	& $*$	& $-1$
	& $ S^+_j S^+_{j+1} S^-_{j+2} -  S^-_j S^+_{j+1} S^+_{j+2}$
	& $\frac{1}{i}[\bar{\partial} O_{1,1}- \partial O_{-1,1}]$ \\

$0$ 	& $0$	& $-1$	& $-1$
	& $ S^z_j(S^+_{j+1}S^-_{j+2}+S^-_{j+1}S^+_{j+2})$
	& $\frac{1}{i}[O_{2,0} -O_{-2,0}]$ 	\\

\medskip

	&	&	&
	&$ \quad - (S^+_j S^-_{j+1}+S^-_j S^+_{j+1})S^z_{j+2}$	&	\\

$0$ 	& $0$	& $1$	& $1$	& a part of the Hamiltonian
	& $O_{2,0}+O_{-2,0}$	\\
\hline

\end{tabular}

\end{table}

\end{document}